\documentclass[12pt]{spieman}  
\usepackage{lineno}
\usepackage{amsmath,amsfonts,amssymb}
\usepackage[dvipsnames]{xcolor}
\usepackage{graphicx}
\usepackage{setspace}
\usepackage{tocloft}
\newcommand{\sw}[1]{\texttt{#1}}
\title{Characterization of a deep-depletion 4K x 4K CCD Detector System designed for ADFOSC}

\author[1, 2]{Dimple* }
\author[1]{T. S. Kumar}
\author[1]{A. Omar}
\author[1]{K. Misra}
\affil[1]{Aryabhatta Research Institute of observational sciences, Manora Peak, Nainital 263001, India.}
\affil[2] {Department of Physics, Deen Dayal Upadhyaya Gorakhpur University, Gorakhpur-273009, India.} 

\cftpagenumbersoff{figure}
\cftpagenumbersoff{table} 
\begin{document} 
\maketitle

\begin{abstract}
We present the characterization of the CCD system developed for the ADFOSC instrument on the 3.6m Devasthal Optical Telescope (DOT). We describe various experiments performed to tune the CCD controller parameters to obtain optimum performance in single and four-port readout modes. Different methodologies employed for characterizing the performance parameters of the CCD, including bias stability, noise, defects, linearity, and gain, are described here. The CCD has grade-0 characteristics at temperatures close to its nominal operating temperature of $-120^\circ$C. The overall system is linear with a regression coefficient of 0.9999, readout noise of 6 electrons, and a gain value close to unity. We demonstrate a method to calculate the dark signal using the gradient in the bias frames at lower temperatures. Using the optimized setting, we verify the performance of the CCD detector system on-sky using the ADFOSC instrument mounted on the 3.6m DOT. Some science targets were observed to evaluate the detector's performance in both imaging and spectroscopic modes.
\end{abstract}


{\noindent \footnotesize\textbf{*}Dimple,  \linkable{dimplepanchal96@gmail.com} }

\begin{spacing}{2}   

\section{Introduction}
\label{sect:intro}  
The 3.6m DOT was commissioned at the Devasthal observatory of Aryabhatta Research Institute of observational sciencES (ARIES), Nainital (India) \cite{2018BSRSL..87...29K}. The Devasthal Observatory is situated in the Himalayan regions of Uttarakhand at $\sim 2450$ meter above the mean sea level with geographical coordinates of $29^{\circ}.360$ N, $79^{\circ}.690$ E. This location lies in the middle of the $180^{\circ}$-wide longitude-gap between the Canary Islands ($\sim 20^{\circ}$ W) and Eastern Australia ($\sim 160^{\circ}$ E), making it suitable for observations of time-critical astronomical events due to the availability of several moderate aperture telescopes. The DOT uses a $f/9$ Ritchey-Chr$^{'}$etien (RC) system supported on an alt-azimuth mount \cite{Sagar_2019, 2012SPIE.8444E..1VN}. The aperture of this telescope is appropriate for medium-resolution spectroscopy and observations of faint sources. A low-dispersion spectrograph-cum-imager, ARIES Devasthal Faint Object Spectrograph (ADFOSC), has been developed in ARIES for spectroscopy and imaging of the celestial objects \cite{2019arXiv190205857O}. The spectrograph uses a fixed focal reducer, which converts the incoming f/9 optical beam from the telescope into a faster $\sim f/4.2$ beam. The spectrograph can be used in both spectroscopic and imaging modes by selecting the instrument's corresponding optical elements (e.g. filters, grism, slit, etc.) with the help of a GUI-based instrument control software \cite{10.1117/12.2233082}. In either mode, a charge-coupled device (CCD) is required to detect and record the data. A CCD detector system/imager has been designed and assembled in ARIES in technical collaboration with the Herzberg Institute of Astrophysics (HIA), Canada.
\begin{figure}
    \centering
    \includegraphics[width=\columnwidth]{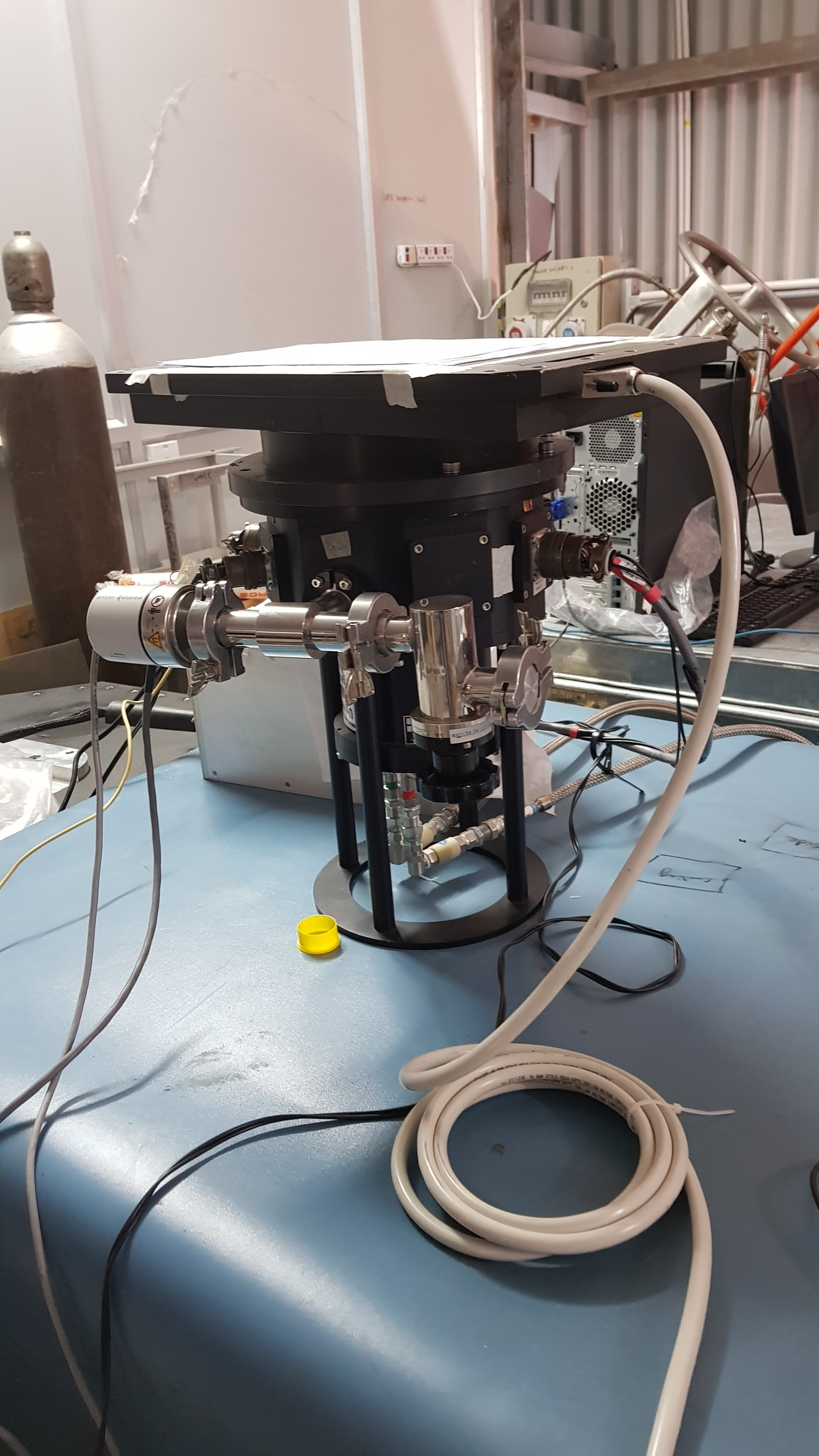}
    \caption{The ADFOSC CCD camera setup in the laboratory. The Camera consists of a CCD detector, a controller, a pressure gauge, a dewar which is cooled using a heat-exchange cryogenic system.}
    \label{fig:Camera}
\end{figure}
\label{ccd_details}

We performed a detailed characterization of the CCD system before commissioning it for scientific observations, both in the laboratory and on the sky. This included estimating parameters like bias level, readout noise, and thermal noise in the dark room. We then performed iterative experiments in the laboratory to optimize the overall system performance and verified the CCD for cosmetic defects. We demonstrate a method to calculate the dark signal of the CCD at different temperatures using the bias frames. As the CCD is developed for the ADFOSC instrument, we also estimated the spectral dispersion on the detector using the lamp spectra. After optimization in the laboratory environment, we performed similar experiments over the night sky on the 3.6m DOT to verify the on-sky performance of the detector system. 

The paper discusses the different methodologies employed for characterizing the performance of the CCD system. The test setup used for performing different tests to optimize the system parameters is detailed in section \ref{characterization}. We also discuss various experiments performed to determine and optimize the CCD characteristics. The integration of the CCD system with the ADFOSC instrument and results of the on-sky tests are discussed in section \ref{sky_verification}. To evaluate the performance of the CCD system on science targets, we observed transient sources during the observing cycle 2020C2 of the 3.6m DOT. The results of the scientific observations are presented in section \ref{performance}.

\section{The CCD detector system}
\label{CCD}
\begin{table}
\centering
\caption{CCD Characteristics}
\begin{tabular}{ll}
\\
\hline
\\
\bf {CCD Characteristic} & \bf {Value} \\
\hline 
\\
CCD & $e2v ~ 231-84$ / Grade-0\\ 
Pixel Size & $15 ~ \mu$m \\ 
Readout Frequency & $160$~kHz\\ 
Operating Temperature & $-120^\circ$C \\ 
Bias level & $1134\pm2.62$~ADU\\ 
Gain & $1.00\pm0.04$ e$^{-}$/ADU \\
Readout Noise & 6~e$^{-}$/pixel \\
Full well capacity & $408$~ke$^{-}$/pixel\\
Saturation level & 65535 ADU \\
\\
\hline
\end{tabular}
\label{tab:ccd_characteristics}
\end{table}
The CCD is a $4096\times4096$ format back-illuminated e2v 231-84 CCD sensor having square pixels of 15-micron size. It is a deep-depleted sensor capable of enhancing the sensitivity toward the longer wavelengths ($ \sim 700-1000$ nm ) of the optical spectrum. The CCD has an imaging area of $61.4$ $\times$ $61.4 \ \rm mm^{2}$, providing a Field of View (FoV) of $\sim13.6\times13.6$ arcmin$^2$ and a spectral dispersion in the range $0.1-0.2$ nm/pixel. The CCD has four readout ports (0, 1, 2, 3). The $\sim16$ million pixels can be read from any of the four amplifiers or through four amplifiers simultaneously. The four-port readout decreases the readout time by a factor of four. However, it requires additional processing to match the bias levels of the four quadrants. Since the readout noise would differ for the four different amplifiers, each quadrant's respective signal-to-noise ratio (SNR) would also be different. In the case of the ADFOSC instrument, we have implemented a single port readout via port-0, which provides the lowest readout noise. The readout frequency is fixed at $\sim 160$ kHz, providing a readout time of $\sim 104$ sec. A Bonn shutter is mounted at the camera entrance with an aperture size of $100$ mm $\times 100$ mm. The shutter employs servo motors for fast operation and offers uniform exposure at the detector plane. Using this shutter, a minimum exposure time of $\sim 1$ ms is possible with an uncertainty of 300 $\rm \mu$s. The detector system consists of the CCD detector, a clock-shaping fan-out circuit board, and a generic Astronomical Research Cameras (ARC) controller\footnote{\url{http://astro-cam.com}} to generate the suitable clock and bias voltages for the detector. The controller hardware has two video cards for reading four ports with 16-bit Analog to digital converter (ADC) units to interface and digitize the four channels of the CCD. Additionally, different bin settings are implemented to read the image in different binning patterns for photometry and spectroscopy.

The CCD sensor is cooled to $-120^\circ$C to minimize the dark signal. The CCD dewar is evacuated for several hours using an oil-free turbo molecular vacuum pump before deep cooling. The dewar is equipped with a Pirani vacuum gauge for monitoring its vacuum pressure. Once the vacuum reaches below $ \sim 1$ x $10^{-3} $ Torr, the closed-cycle (Joule-Thomson) cryogenic heat-exchange system supplied by Brooks Automation, USA, is switched on. The overall process of cryo-cooling the CCD from an ambient temperature of $25^\circ$C to $-120^\circ$C takes between 4 to 5 hours. This temperature is stabilized and held constant within $0.01^\circ$C using a Lakeshore 335 Proportional Integral Derivative (PID) temperature controller\footnote{\url{http://irtfweb.ifa.hawaii.edu/~s2/software/gpib-eth-ls335/335_Manual.pdf}}. For this purpose, a small heater and a temperature sensor are mounted on the cold plate below the sensor. A charcoal-filled getter is used to absorb outgassing inside the dewar. The charcoal gets activated to absorb gases at cryogenic temperatures and helps attain a high vacuum. An ultimate vacuum of $ \sim 3$ x $10^{-7} $ Torr is usually attained with this system at $-120^\circ$C.

\section{Characterization of the CCD system}
\label{characterization}
Detailed characterization of the CCD includes the estimation of bias level, bias stability, readout noise (RN), gain, defects, linearity, and saturation level of the CCD. This section describes the laboratory-based test setup and the experiments performed to determine these parameters. We tested the CCD performance using all four ports, numbered zero to three. However, the read noise is found to be the lowest for port-0; hence single port readout mode using port-0 is currently being used for acquiring the scientific data. The paper focuses on characterizing parameters for port-0 of the CCD system. 

\subsection{Test setup and Data Acquisition}
\label{set-up}
We set up the CCD system on an electrostatic discharge (ESD) safe, dark room of the ARIES optics laboratory for performing the experiments. A light-emitting diode (LED) operated at a constant regulated voltage was used as an illumination source for the experiment. We covered the CCD window with an Aluminium plate with a pinhole for the light to enter. We fixed the source on the pinhole to avoid any fluctuations in light intensity due to any change in the source's position. The sub-systems, namely, the temperature controller, cryogenic pump, pressure gauge, etc., were carefully grounded to a common point to avoid noise entering from ground loops. The entire system was again reconfigured when the ADFOSC was mounted on the 3.6m DOT for sky tests. We acquired the data using the \sw{Owl}\footnote{\url{http://www.astro-cam.com/Gen3Software.php}} software provided along with the ARC controller. The software offers different dialog boxes to control the controller parameters, including gain, readout, binning, etc., and saves the acquired images in standard Flexible Image Transport System (FITS) format. Different modules of \sw{Python}\cite{van1995python} like \sw{Astropy}\cite{astropy:2013,astropy:2018,astropy:2022} and \sw{ccdproc}\cite{matt_craig_2017_1069648} were used to process the FITS image files.

\subsection{Bias level and readout noise}
\label{sec:bias_level}
\begin{figure}
\includegraphics[width=\columnwidth]{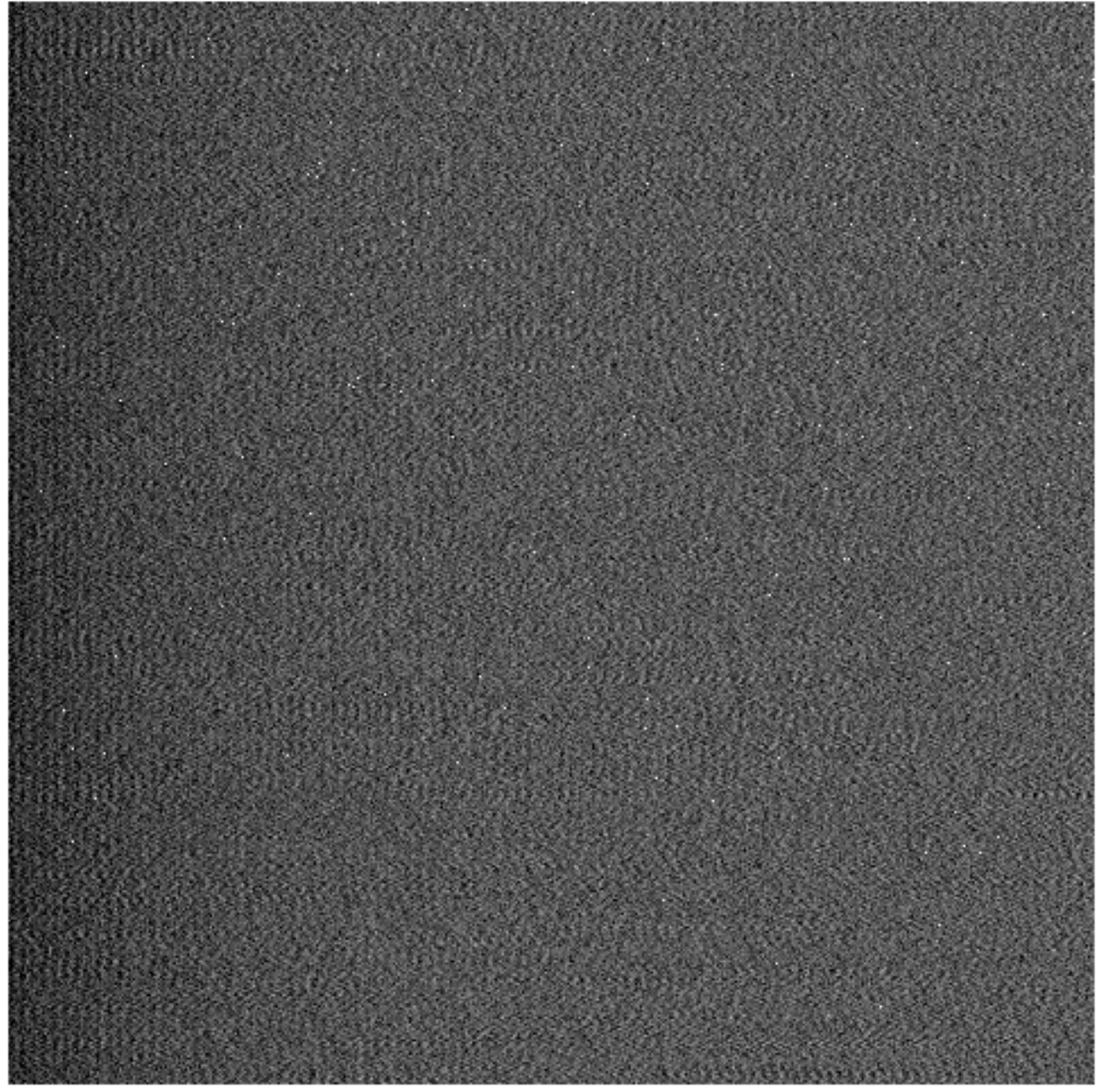}
\caption{Master bias of the CCD created using median combining 50 bias frames.}
\label{bias_frame}
\end{figure}

A positive offset is generally provided to the CCD electronics to avoid negative counts in the output of the CCD. The mean offset value, or the bias value, is optimized in a way that it is large enough to avoid the non-linear regime of the CCD amplifiers but not too large to reduce the dynamic range of the CCD. We set the bias level slightly above thousand counts to balance the above factors.

\begin{figure}
\includegraphics[width=\columnwidth]{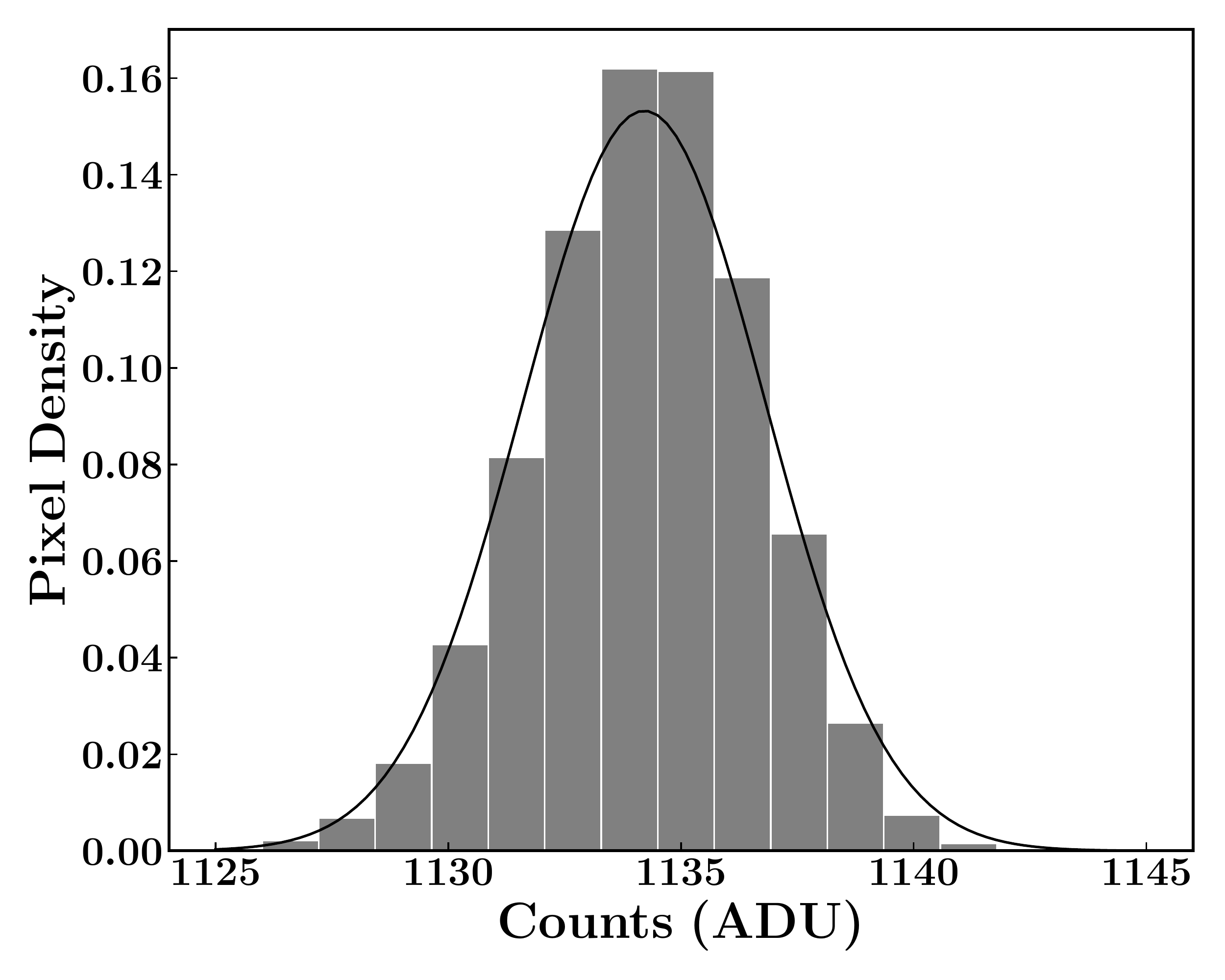}
\caption{Histogram of the master bias frame with a mean value of $1134.01\pm2.62$ counts.}
\label{bias}
\end{figure}

We estimated the bias level of the CCD using the bias frames, which are the CCD images with a zero-second exposure. Several bias frames were acquired, and we used fifty of them to generate a master bias frame by taking their median value. We did this using \sw{Mediancombine} task of \sw{ccdproc} software module. Fig. \ref{bias_frame} shows the median bias frame, and the corresponding histogram is shown in Fig. \ref{bias}. The width of the distribution represents the RN of the CCD, which is the number of electrons introduced by the readout electronics while reading out each pixel. We estimated the RN using the Janesick method (equation \ref{eq:RNEQ})\cite{2001sccd.book.....J}. We created a difference image using two bias images and estimated its standard deviation ($\sigma_{Bias1-Bias2}$).
The explanation of the Gain value used in equation \ref{eq:RNEQ} is provided in section \ref{gain}. We found the RN value to be 6.20 analog to digital units (ADU) or 6.20 $e^{-1}$ for the gain value of $\sim$ 1. However, the achieved noise is more than twice the value in the e2v datasheet at 160kHz. We observed some interference patterns in the image. These patterns are likely to be responsible for this increased noise. The probable cause could be ground loops outside and inside the dewar, length of cables and imperfect shielding scheme. These have been controlled to a large extent by iteratively evaluating various schemes like shortening the cables and star connecting the ground points of the auxiliary devices like chiller, pressure gauge, temperature controller etc. Also, grounding permutations were tried with the four-port video cables. Finally, the shielding of the video cables was grounded only at the controller end and left open at the CCD connector end, which resulted in a lower noise floor. There is still scope for improvement as the ground loops inside the dewar have not been evaluated. This evaluation will be attempted later since the CCD has already been commissioned for observations.

\begin{center}
\begin{equation}
RN =  \frac{Gain\times\sigma_{Bias1-Bias2}}{\sqrt{2}} \label{eq:RNEQ},
\end{equation}
\end{center}

\begin{figure}
\includegraphics[width=\columnwidth]{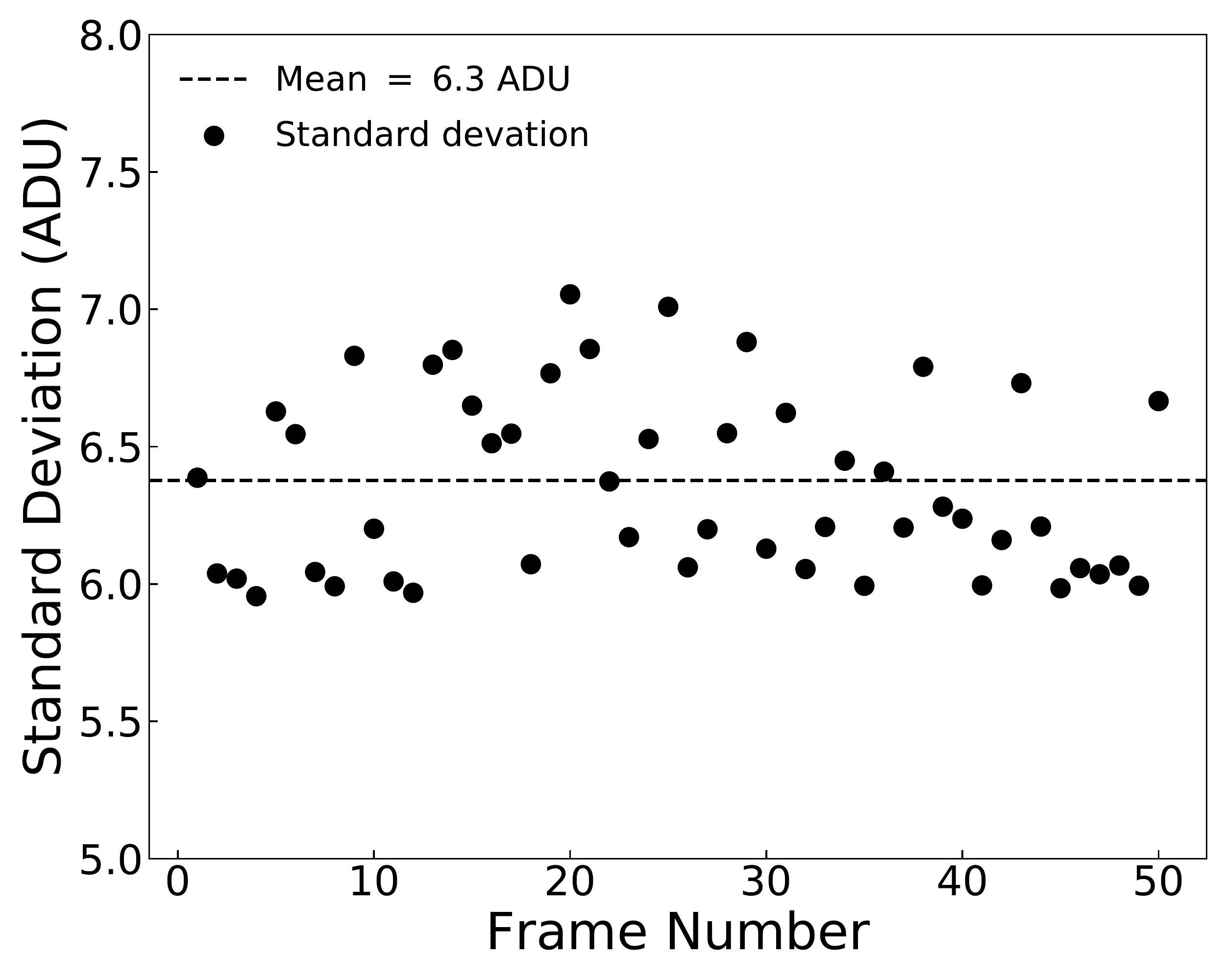}
\caption{A plot showing the variation of the standard deviation of 50 bias frames}
\label{RN}
\end{figure}

We calculated the standard deviation of 50 bias frames to verify this RN value, as shown in Fig. \ref{RN}. The mean of these standard deviation values is 6.38 ADU, which is consistent with the value calculated using equation \ref{eq:RNEQ}.

\subsection{Linearity}
The CCD system should have a linear response to the incident light for scientific observations. However, several factors can introduce non-linearity in CCD performance. The controller clock periods and overlaps should be timed for complete charge transfer during the readout process. Moreover, there should be a delay between this transfer and the correlated double sampling instant to allow the transients to settle to avoid any induced noise or glitch and introduce non-linearity. We verified the signal waveform using a 1 GHz digital oscilloscope to ensure the above before connecting the interface board to the detector. Other factors critical for linearity are bias voltages: voltage output drain (VOD) and voltage reset drain (VRD). The CCD manufacturer provided a range of values for the voltages, VOD from 25 to 31 volts and VRD from 16 to 19 volts. To check the behaviour of the CCD at different voltages, we initially set these voltages near minimum values and iteratively increased these voltages within this range. We rejected some of the voltage combinations that provided very low-bias levels. For other combinations, experiments were performed to check the linearity of the CCD. 

We used an LED source (section \ref{set-up}) to illuminate the CCD and acquired images with an incremental increase in exposure time. We obtained a pair of images for each exposure time to detect any variation in the source intensity. We noticed that the counts are identical for the pair of images.

\begin{figure}
\centering
\includegraphics[width=\columnwidth]{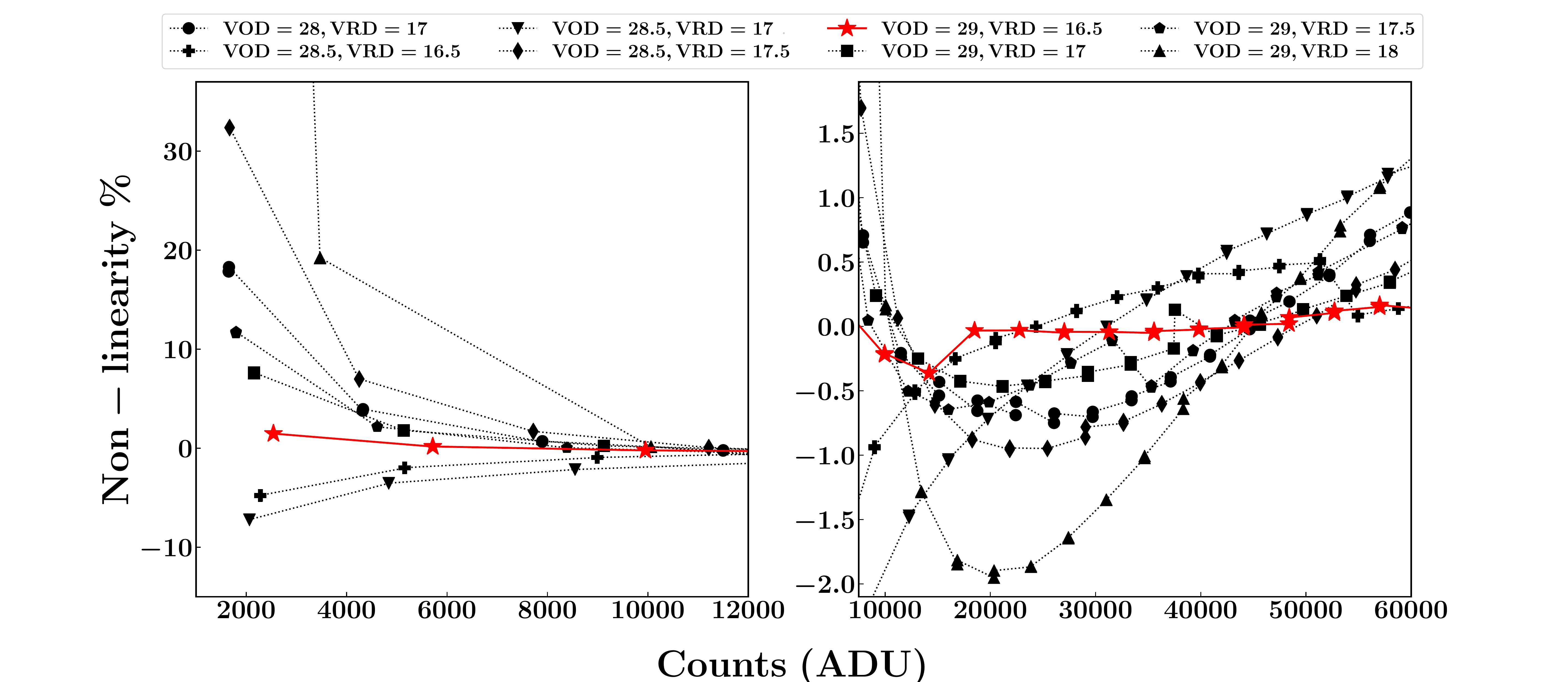}
\caption{Non-linearity curves at different operating voltages in the lower count regime (left panel) and in the higher count region (right panel). Non-linearity is the minimum for a combination of VOD=29 volts and VRD=16.5 volts.}
\label{linearity_ccd}
\end{figure}

We estimated the non-linearity (the relative difference between our measurements and the best-fit linear curves) by fitting a linear function to the variation of mean counts with exposure time for each voltage combination. We used the \sw{linregress} function from the \sw{stats} library under \sw{Python} for this purpose. Fig. \ref{linearity_ccd} shows the non-linearity curves for various combinations of VOD and VRD in different count regions. For most voltage combinations, the non-linearity is negligible in the higher count regime. The non-linearity, however, shows up in the lower count regime and is significant for certain voltages. For a combination of VOD = 29 volts and VRD = 16.5 volts, the non-linearity is the lowest. For this voltage combination, the value of the regression coefficient ($R^2$) is $0.9999$, which is almost equal to unity (see Fig. \ref{linearity}). We considered this voltage combination as the optimum value for the CCD system. 

\subsection{Saturation level}
\begin{figure}
\centering
\includegraphics[width=\columnwidth]{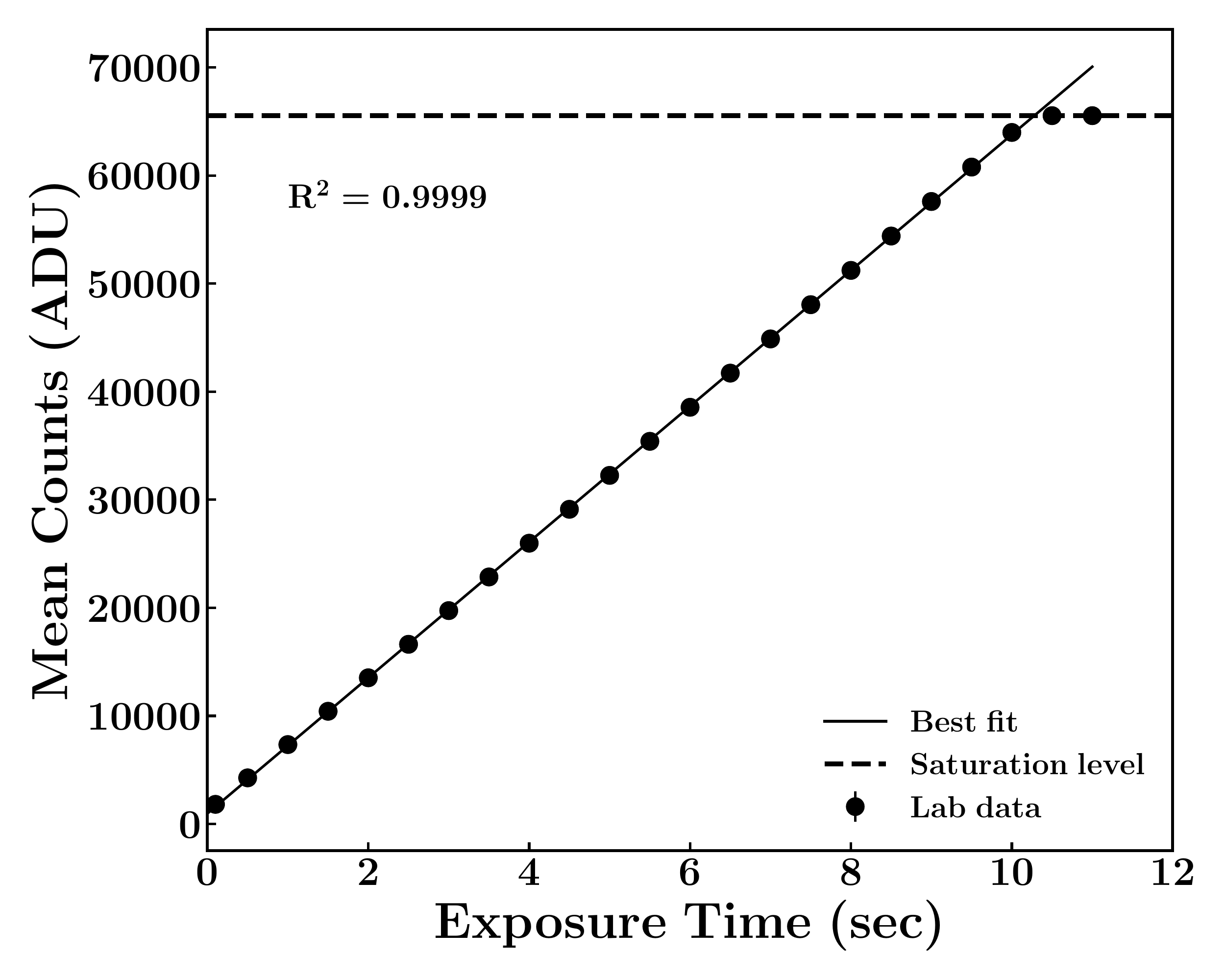}
\caption{Linearity curve at VOD = 29 volts and VRD = 16.5 volts. A linear fit to the data gives the regression coefficient as 0.9999. The horizontal dashed line indicates the saturation level.}
\label{linearity}
\end{figure}

The maximum capacity of a CCD pixel to store the photo-electrons is its full well capacity, beyond which the pixels saturate. Since the available 16-bit ADC of the controller saturates at a value of 65535, the controller's gain setting helps to select the dynamic range. As we are interested in accurate photometry of faint objects, a gain of unity is selected, constraining the saturation point to 65535. The selection of system gain is discussed in section \ref{gain}. Illumination of the CCD until its saturation point limited the counts to 65535, the saturation point of the 16-bit ADC. It is demonstrated and shown in Fig. \ref{linearity} where a bright source illuminates the detector, and the ADC saturates at 65535 counts. If the science cases demand the utilization of full well capacity, the user can select a gain setting close to 3 or higher electrons per ADU. 

\subsection{The system gain}
\label{gain}
The gain of a CCD system is defined in terms of ADU, which corresponds to the number of electrons assigned to one digital unit in the output image. The available gain values are 1, 2, 5, and 10 {e$^{-}/$ADU} in the controller. The gain values can be selected using the software at runtime. The saturation level of the CCD should correspond to the saturation level of the ADC to utilize the full well capacity. Since the full well capacity of the CCD is $408$~ke$^{-}$ (as mentioned in the result sheet of the supplied detector),  a gain of 10 is suitable to match the saturation levels. However, for the detection of photon-limited faint objects, a gain of 1 is implemented using the controller parameters.

The electronic gain of the CCD system is the product of gain values introduced by each stage of the readout electronics. The inherent gain of the CCD, defined by the output capacitor, is 7 {$\mu$ V/e$^{-}$}. A series of Op-amp stages within the controller further amplify this gain. Initially, it is preamplified with a gain of 4 and passed through a gain selection stage, offering a range of gain values: 1, 2, 4.75, and 9.5. A bias adjustment stage after the integrator provides a gain of 2. Hence, an amplification of 56 {$\mu$ V/e$^{-}$} is obtained with these four stages. Since the 16-bit ADC operating at a reference voltage of 10 V provides a bin size of 152.588 {$\mu$ V/ADU}, the integration of the Op-amp integrator is adjusted to provide an additional gain factor of 2.725 to achieve the desired system gain of 1 {e$^{-}/$ADU}. Since the integrator time can only be adjusted in increments of 40 ns, the closest possible value of 0.998 {e$^{-}/$ADU} is set.

We experimentally verified this gain setting using the Janesick method \cite{2001sccd.book.....J} given by equation \ref{eq:gainEQ} where ($S$) is the mean of the signal acquired by the CCD, and $\sigma_S^2$ is the variance.

\begin{center}
\begin{equation}
\sigma_S^2 =  \frac{S}{G} + \sigma_R^2  \label{eq:gainEQ},
\end{equation}
\end{center}

\begin{figure}
   \includegraphics[width=\columnwidth]{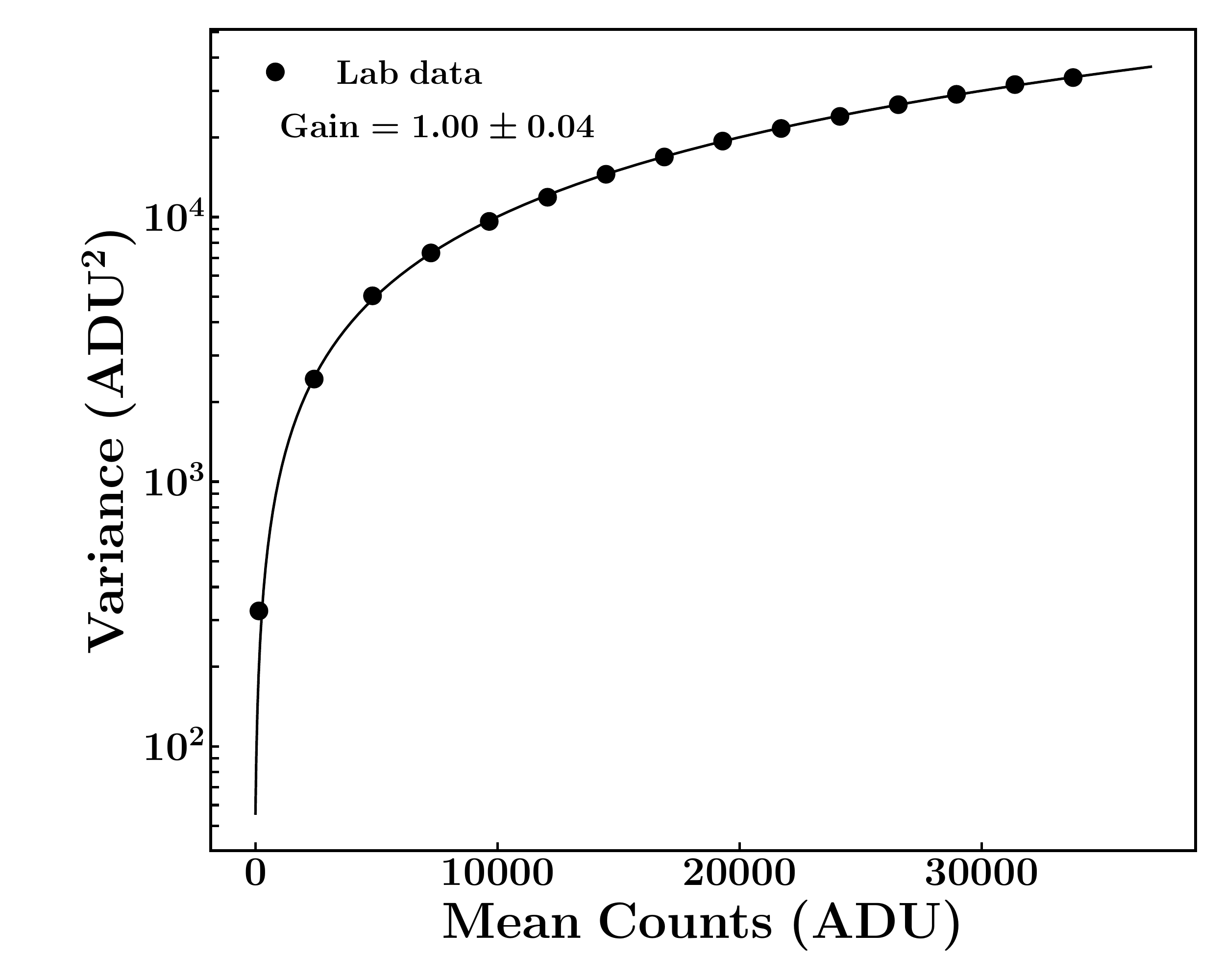}
    \caption{Photon transfer curve (PTC) of the CCD obtained in the laboratory environment. The measured value of the gain is $1.00\pm0.04$ e$^{-}/$ADU.}
    \label{gain_ccd}
\end{figure}

We acquired a pair of images at each exposure and estimated the mean signal after bias subtraction and cosmic-ray removal from the image. Further, these images were normalized by subtracting one image from the other for compensating the flat-field effect. We used the resulting image to estimate the variance of the signal ($\sigma_S^2$). Fig. \ref{gain_ccd} shows the photon transfer curve (PTC) for the CCD. To estimate the gain, the PTC was fitted with a linear function using \sw{linregress}. The estimated gain is 1.00 $\pm$ 0.04 {e$^{-}/$ADU}, which matches the electronic gain value of the system within the errorbar.

\begin{figure*}
\includegraphics[width=8cm,height=8cm]{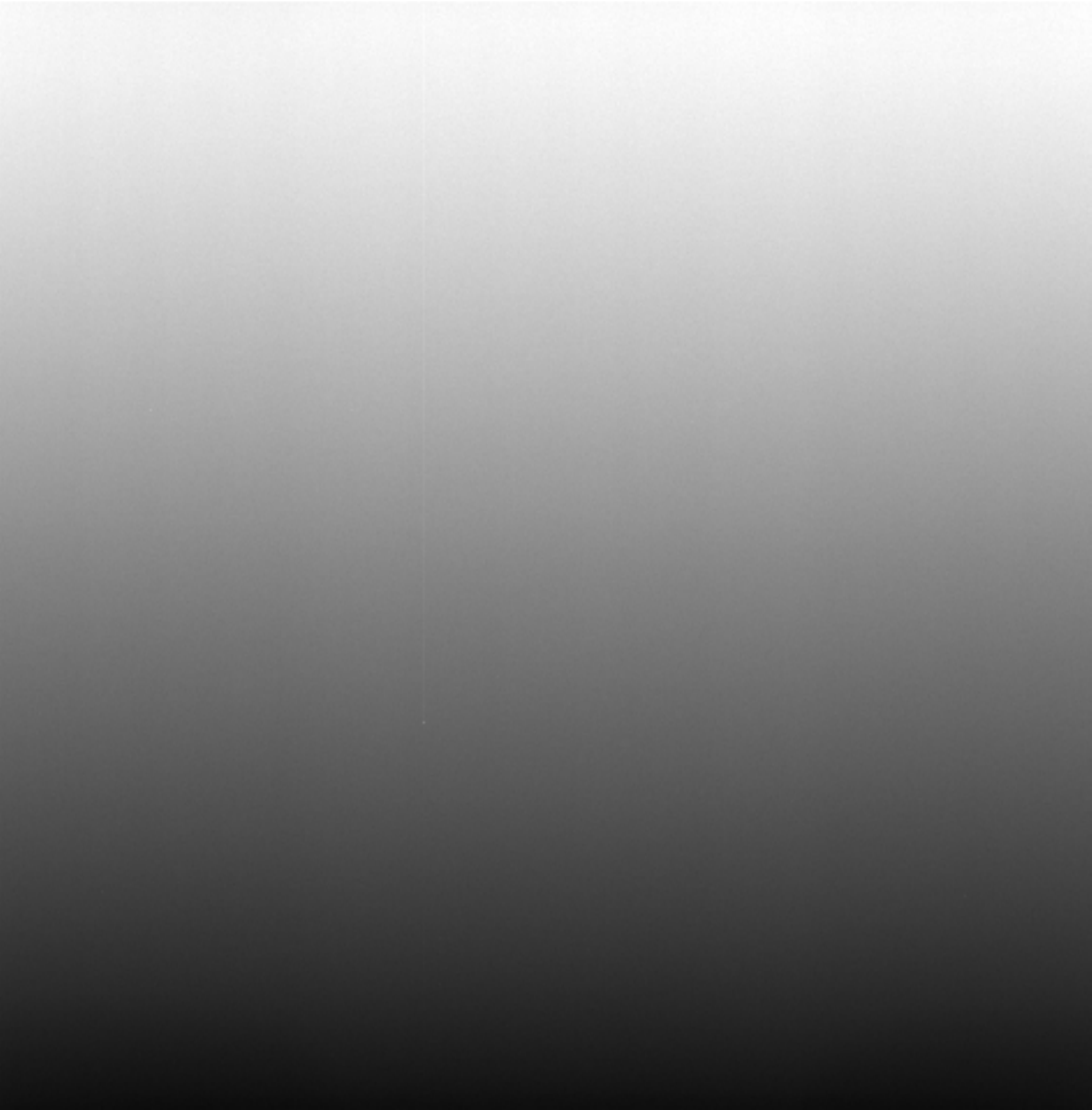}
\includegraphics[scale=0.40]{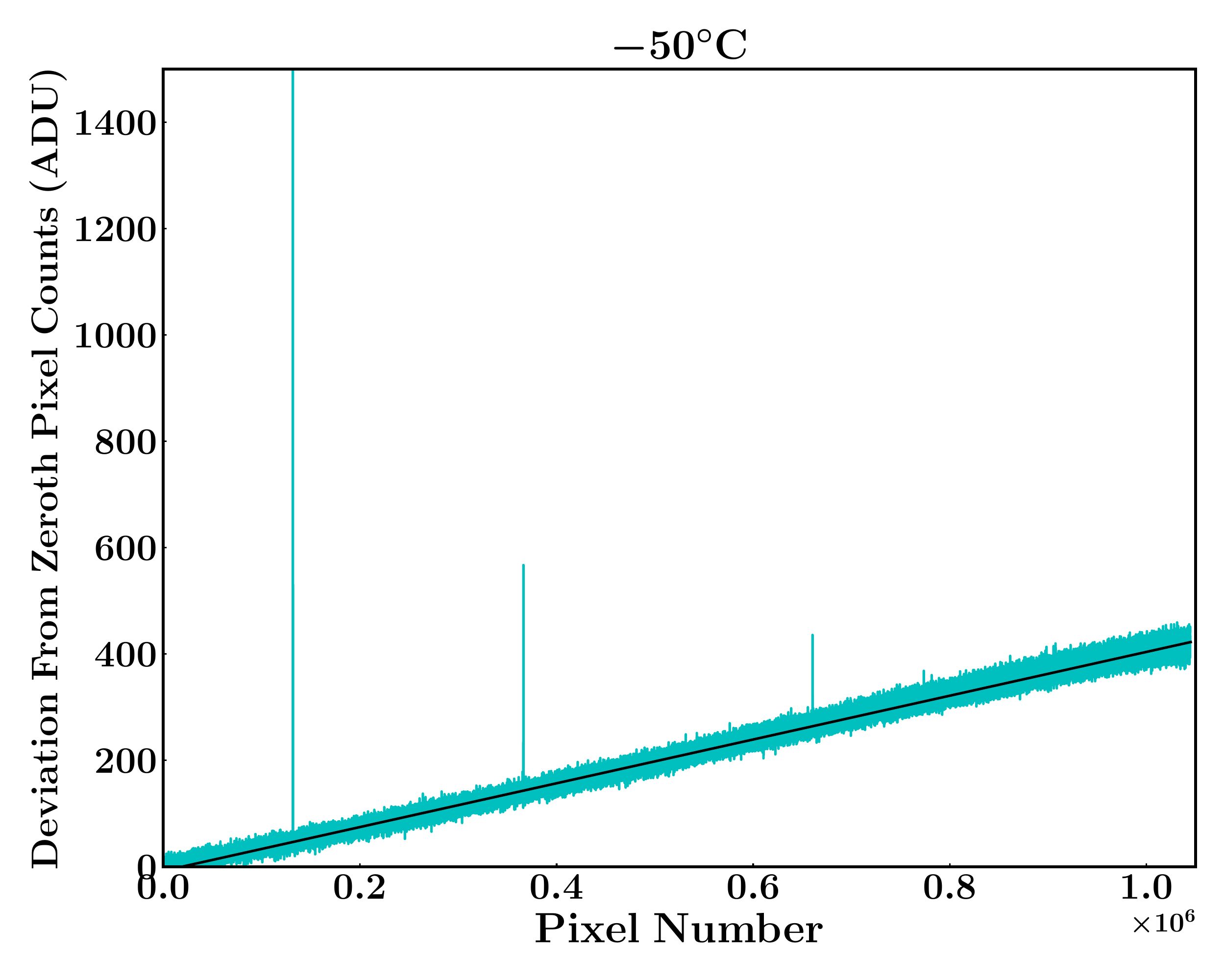}
\caption{Masterbias at $-50^\circ$C and deviation of the counts of all pixels from the zeroth pixel counts. A clear gradient can be seen in the image as the last pixels get more time to generate the dark signal.}
\label{dark}
\end{figure*}

\subsection{Thermal noise}
\label{thermal_noise}
In the CCD detectors, electron charge-density increases exponentially with an increase in temperature due to the thermal generation of electrons. A CCD must be cooled optimally to minimize the dark signal. To determine this optimum temperature, we calculated the dark signal at various temperatures ranging from $-35^\circ$C  to $-120^\circ$C using the bias frames. We acquired several bias frames at different temperatures and generated master bias frames at each temperature. The left panel of Fig. \ref{dark} shows the master bias frame at $-50^\circ$C. A gradient in counts is visible in the master bias due to the finite readout time of the CCD. The zeroth pixel gets the lowest time to generate dark electrons. The last pixel accumulates dark counts over full readout time, hence has the maximum number of thermally generated electrons. If the readout time and the gain are known, then by comparing the counts of the first and the last pixel, we can measure the number of electrons generated per pixel per second. Using this method, we estimated the dark signal at different temperatures. 
\vspace{0.8 cm}
\begin{table}[h!]
\begin{center}

\begin{tabular}{ |c c | c c| }
  
\hline
\begin{tabular}[c]{@{}l@{}}Temp.\\ ($^\circ$C)\end{tabular} & \begin{tabular}[c]{@{}l@{}}Dark Current\\ (e-/pix/sec)\end{tabular} & \begin{tabular}[c]{@{}l@{}}Temp.\\ ($^\circ$C)\end{tabular} & \begin{tabular}[c]{@{}l@{}}Dark Current\\ (e-/pix/sec)\end{tabular} \\ 
\hline
~-35.0 & ~~~$39.89\pm0.017$ & ~-60.0 & ~~~$1.03\pm0.002$\\
~-40.0 & ~~~$21.30\pm0.009$ & ~-65.0 & ~~~$0.42\pm0.002$\\
~-42.5 & ~~~$15.36\pm0.007$ & ~-70.0 & ~~~$0.18\pm0.002$\\ 
~-50.0 & ~~~$5.37\pm0.003$ &  ~-80.0 & ~~~$0.10\pm0.001$\\
 ~-55.0 & ~~~$2.89\pm0.002$ & ~-90.0 & ~~~$0.05\pm0.001$\\
~-57.5 & ~~~$1.87\pm0.002$ & ~-95.0 & ~~~$0.01\pm0.002$\\
\hline
\end{tabular}
\vspace{0.5cm}
\caption{Dark current at different temperatures as estimated using the gradient of bias frames.}
    \label{tab:dark}
\end{center}
\end{table}

The right panel of Fig. \ref{dark} shows the deviation of counts in each pixel from the zeroth counts. The farther the pixel number is from the readout port, the larger the dark count and the larger the deviation from the zeroth counts. To determine the slope of this gradient, we fitted a polynomial in counts vs. pixel number data using the \sw{polyfit} function of \sw{Python}. It is seen that a linear function provides the best fit, as shown in the left panel of Fig. \ref{dark}. We used the slope to calculate the difference in counts between the first and the last pixel. We divided this difference by the total readout time to obtain dark counts generated per second. Since the bias frames were acquired in $4\times4$ binning, it was further scaled by a factor of 16 after subtracting the RN. Below $-100^\circ$C temperature, the thermal noise becomes less than RN; hence, we could estimate the dark signal values up to $-95^\circ$C. The dark signal values at different temperatures are listed in table \ref{tab:dark}.

\begin{figure}
\includegraphics[width=\columnwidth]{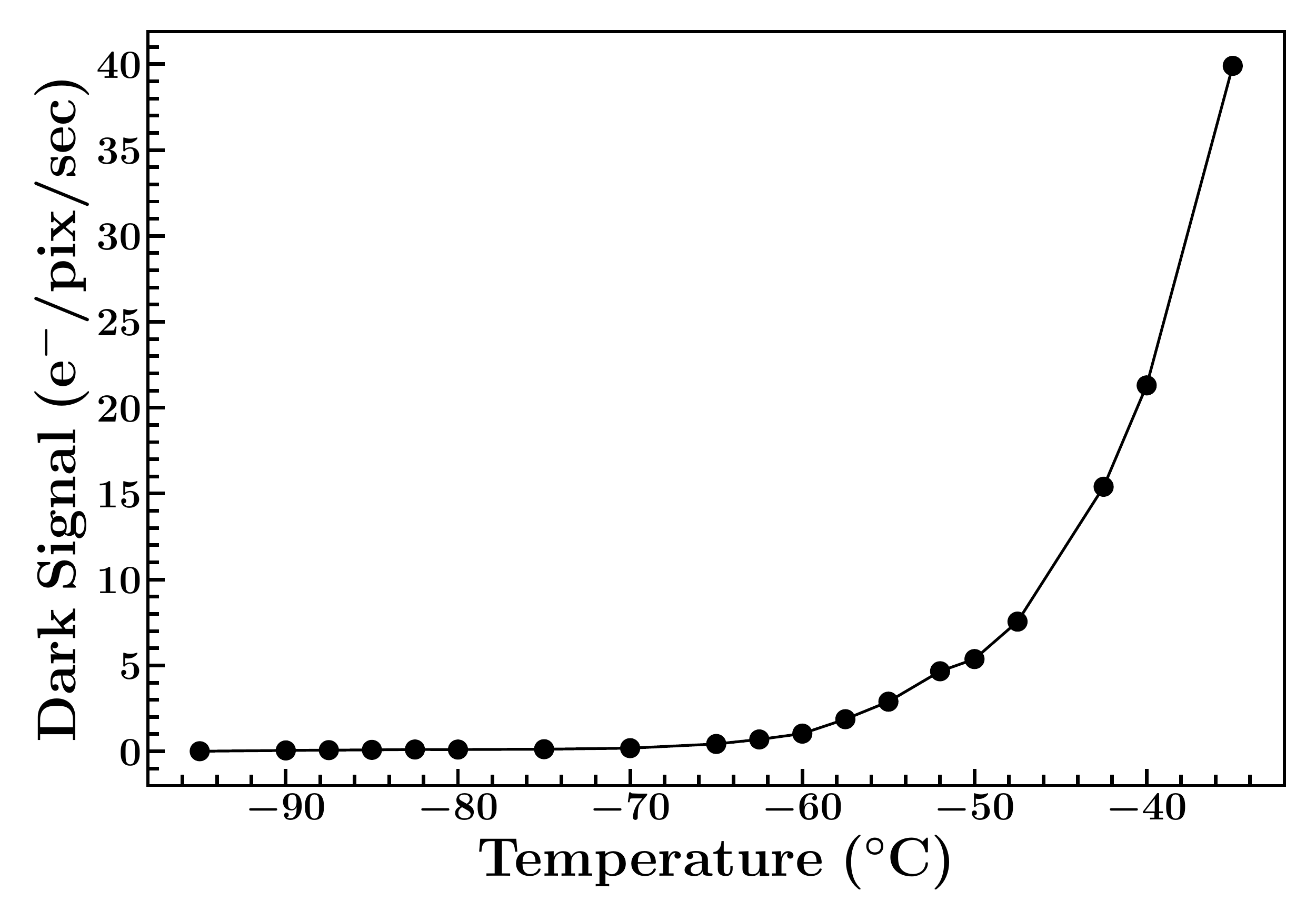}
\caption{Variation of dark signal with temperature. The dark signal is negligible below $-80^\circ$C.} 
\label{dark_signal}
\end{figure}

As shown in Fig. \ref{dark_signal}, the dark signal varies exponentially with temperature. Below -80°C, the dark signal is negligible, suggesting that the CCD can be used below this temperature with minimal thermal noise.

\subsection{CCD defects}
\label{ccd_defects}
\begin{figure}
\includegraphics[width=\columnwidth]{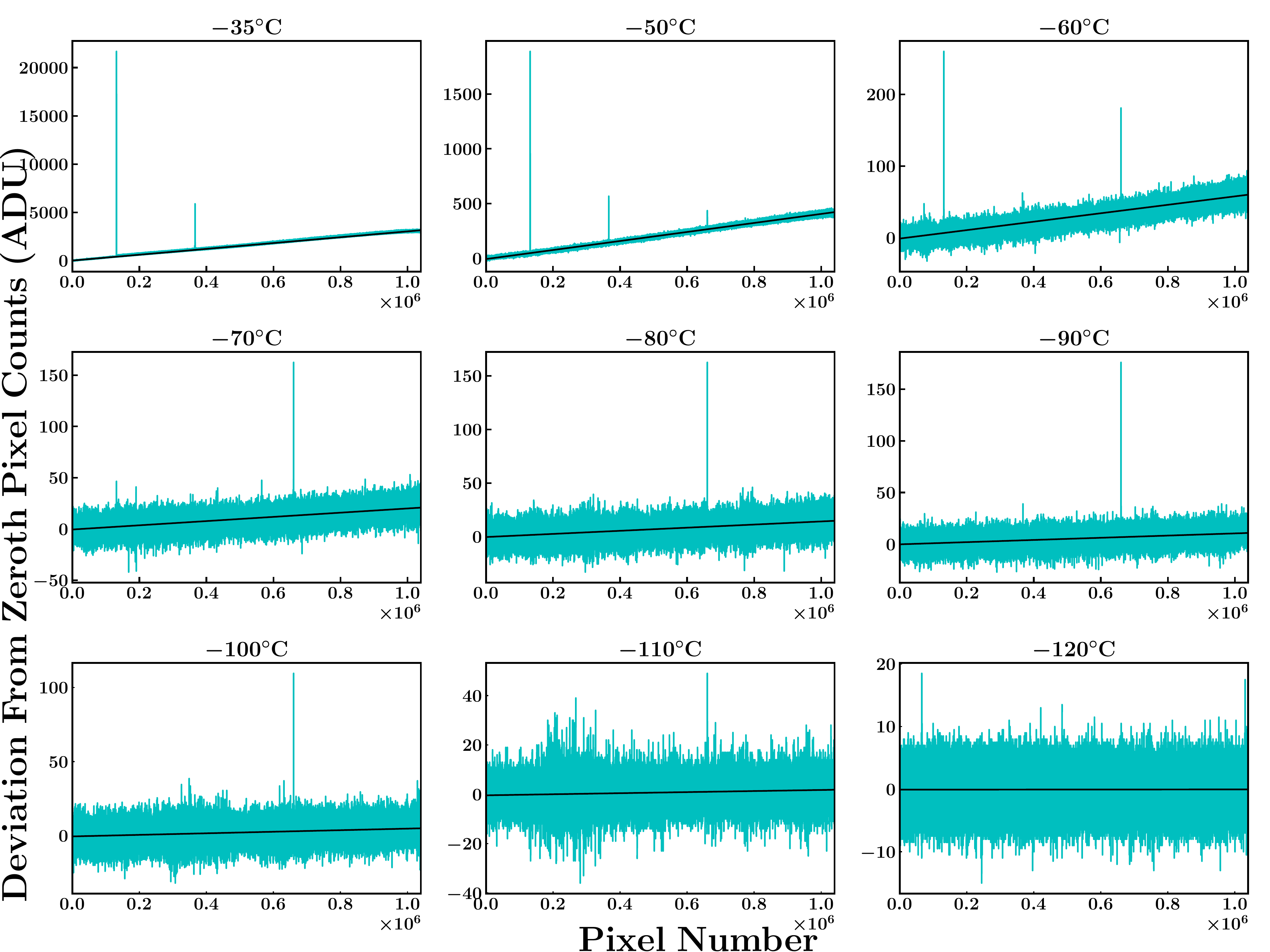}
\caption{The response of CCD pixels at different temperatures is shown in the figure. The deviation of the counts from the zeroth counts is fitted with a polynomial (black line). The dark signal is calculated from the best fit. There are a few pixels which are behaving differently at higher temperatures. At lower temperatures, the CCD behaves as grade-0 CCD.}
\label{defects} 
\end{figure}

The CCD may have some pixels that might not respond to light optimally due to defects in the CCD structure. These can be point defects, hot defects, or dark/dead pixels. We are employing a grade-0 CCD detector (the CCD detector with minimum possible defects) as mentioned by the manufacturer. To check the CCD for point and column defects, we examined the response of all the pixels at different temperatures. Fig. \ref{defects} shows the deviation of counts from the mean bias counts for each pixel of the CCD operated at different temperatures (the method to obtain these plots is described in section \ref{thermal_noise}). Some pixels are seen to behave differently at higher temperatures exhibiting high counts. Though they appear to be hot pixels, the counts are found to decrease with decreasing temperature. Eventually, below $-110^\circ$C, the CCD acts as a nominal grade-0 CCD without any point and column defects.

\section{On-sky verification}
\label{sky_verification}
After optimizing the performance of the CCD system in the laboratory, we verified the on-sky performance. The CCD was integrated with the ADFOSC instrument and mounted on the axial port of 3.6m DOT. This section describes the estimation of gain, linearity, bias level, and bias stability using on sky observation with the instrument.

\subsection{Bias Stability}
We calculated the bias level using the methodology described in section 3. The mean bias level equal to $1133.85\pm2.48$ matches the laboratory estimated value, i.e. $1134.01\pm2.62$. Since fluctuation in bias level can introduce errors in photometric estimates, we acquired and examined several bias frames to check the stability of the bias. Fig. 4 shows the variation of the mean bias level for 30 different nights spread across an observing cycle of three months. The mean bias level fluctuates within a fraction of a count, ensuring bias stability. 
\begin{figure}
\includegraphics[width=\columnwidth]{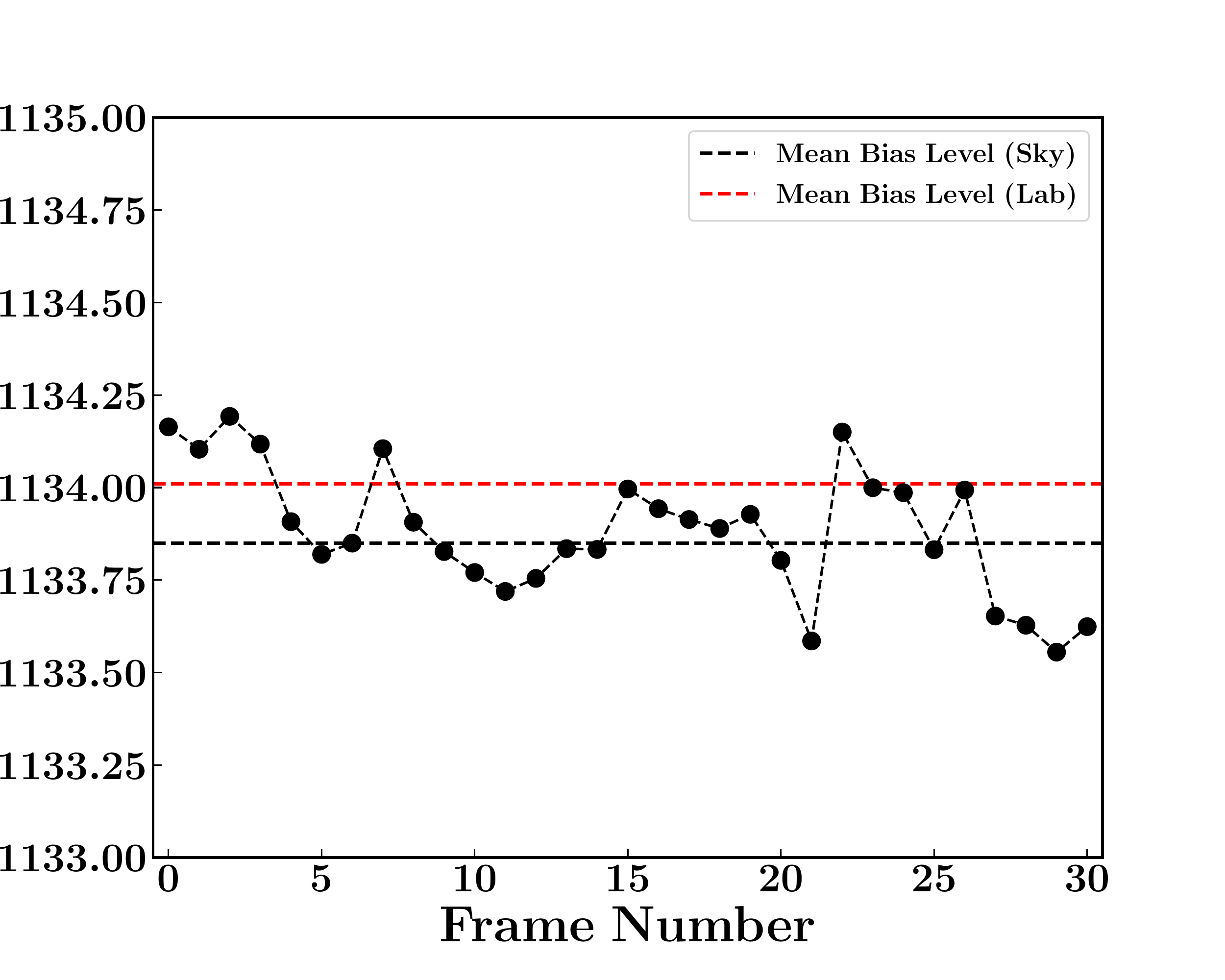}
\caption{Variation within the mean counts of bias frames during different nights. The bias level is stable within one count. The red and black dotted lines show the mean bias level estimated in the lab and on-sky.}
\label{bias_stability}
\end{figure}

\subsection{Linearity and gain}
\begin{figure}
    \centering
     \includegraphics[width=\columnwidth]{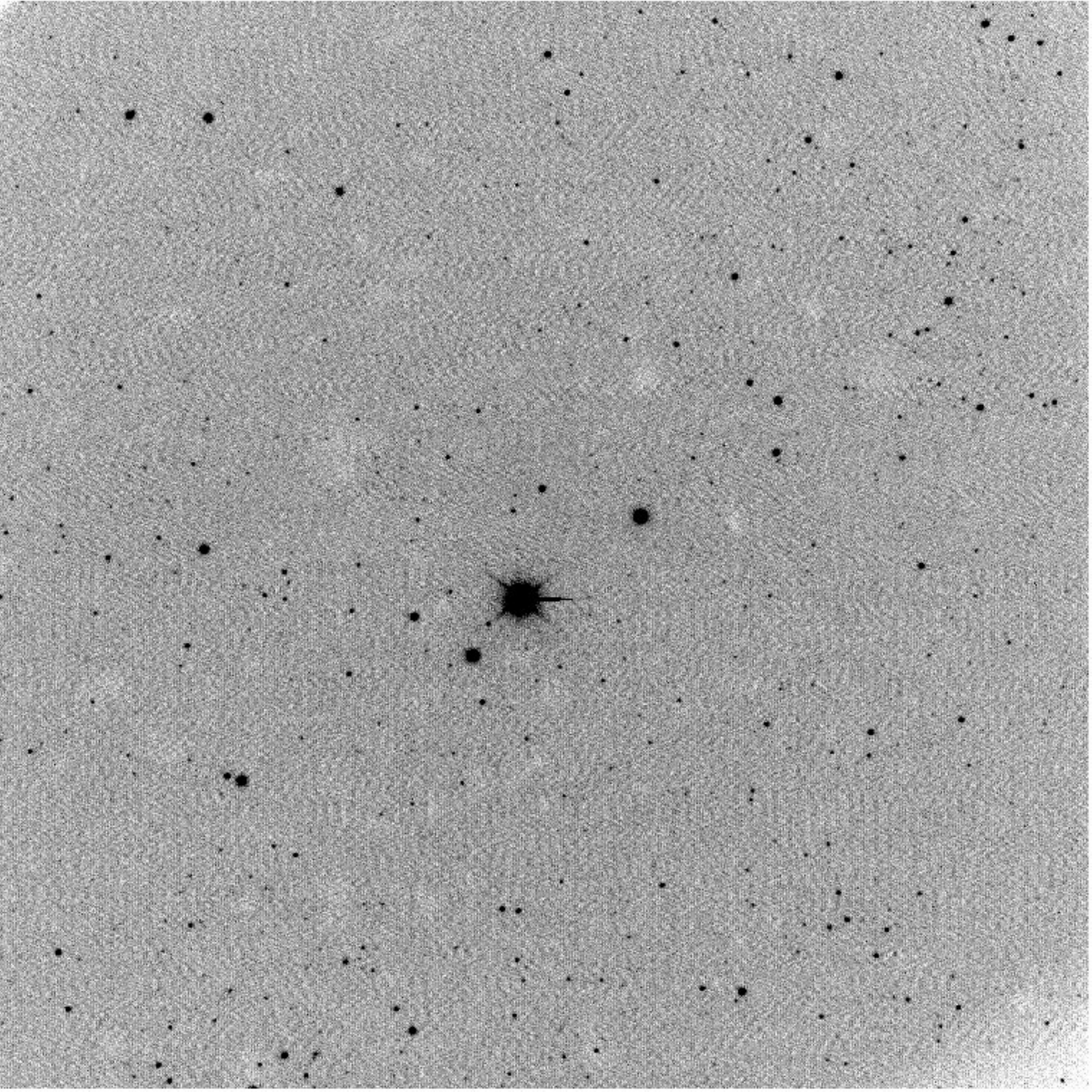}
     \caption{CCD image of the Landolt standard field SA110. The standard stars are between 10 to 16 mag in V-band.}
    \label{SA110} 
\end{figure}
We chose the standard field available at the zenith at the time of observations to validate the linearity and gain of the CCD. Multiple images of Landolt standard field SA110 \cite{1992AJ....104..340L} were acquired in r-band with an exposure time ranging from 5 sec to 100 sec. Before using the images for characterization purposes, we pre-processed the images with basic steps of bias subtraction, flat correction and cosmic-ray removal using the \sw{ccdproc}. Fig. \ref{SA110} shows the pre-processed CCD image of the field SA110, which contains both bright and faint stars (with magnitudes ranging from 10 to 16 mag in the V-band).

\begin{figure}
    \centering
    \includegraphics[width=\columnwidth]{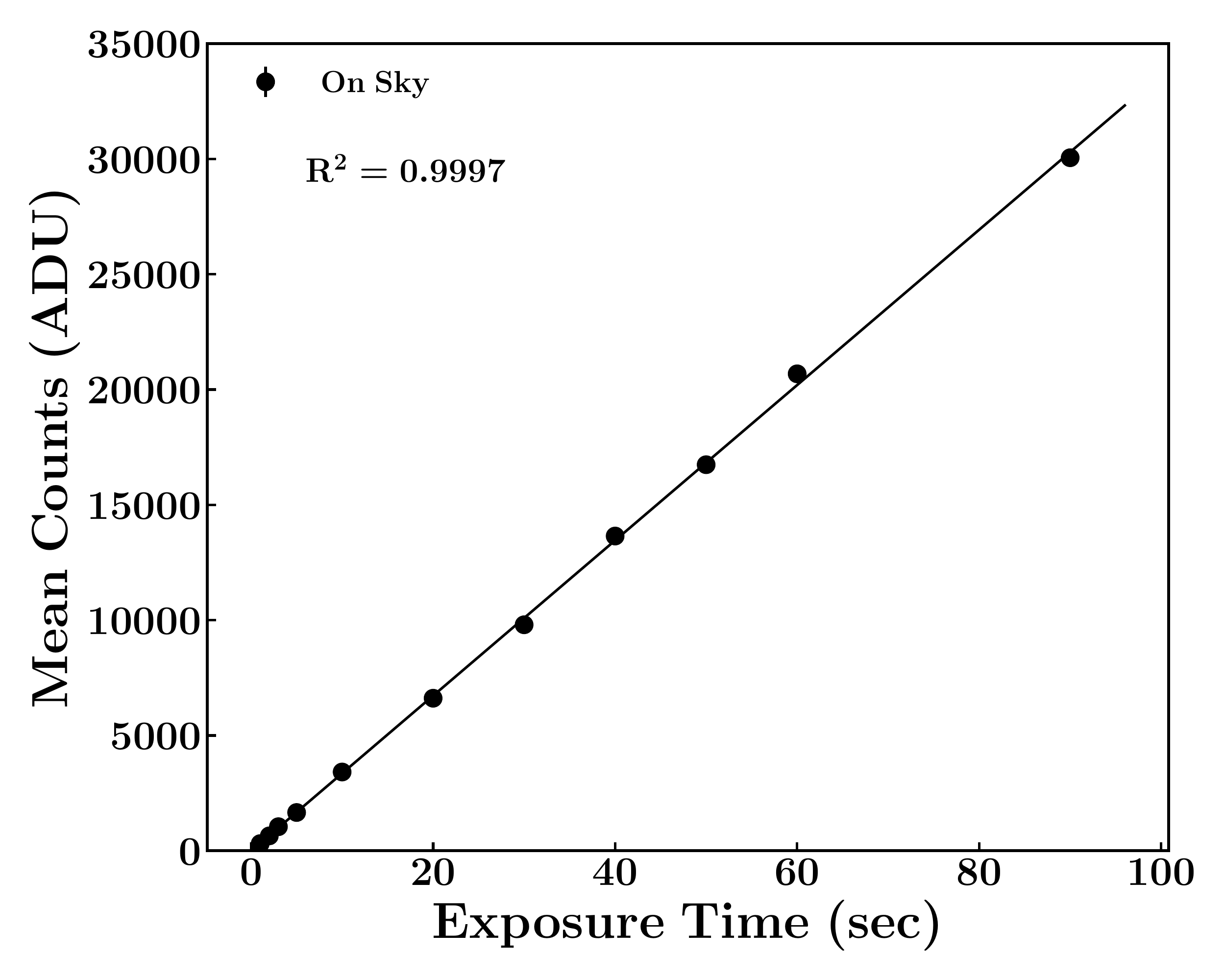}
    \caption{Variation of mean counts with exposure time from on-sky experiments. The black line represents the best linear fit with a regression coefficient of 0.9997. }
    \label{on_sky_linearity}
\end{figure}

We used the faint stars to check the linearity in the lower count region and the bright stars to estimate the saturation level of the CCD. Fig. \ref{on_sky_linearity} shows CCD linearity with $R^2$ = $0.9997$ and a non-linearity percentage of 0.30. The CCD system is seen to saturate at 65535 counts for a gain of 1 {e$^{-}/$ADU}.

\begin{figure}
    \centering
     \includegraphics[width=\columnwidth]{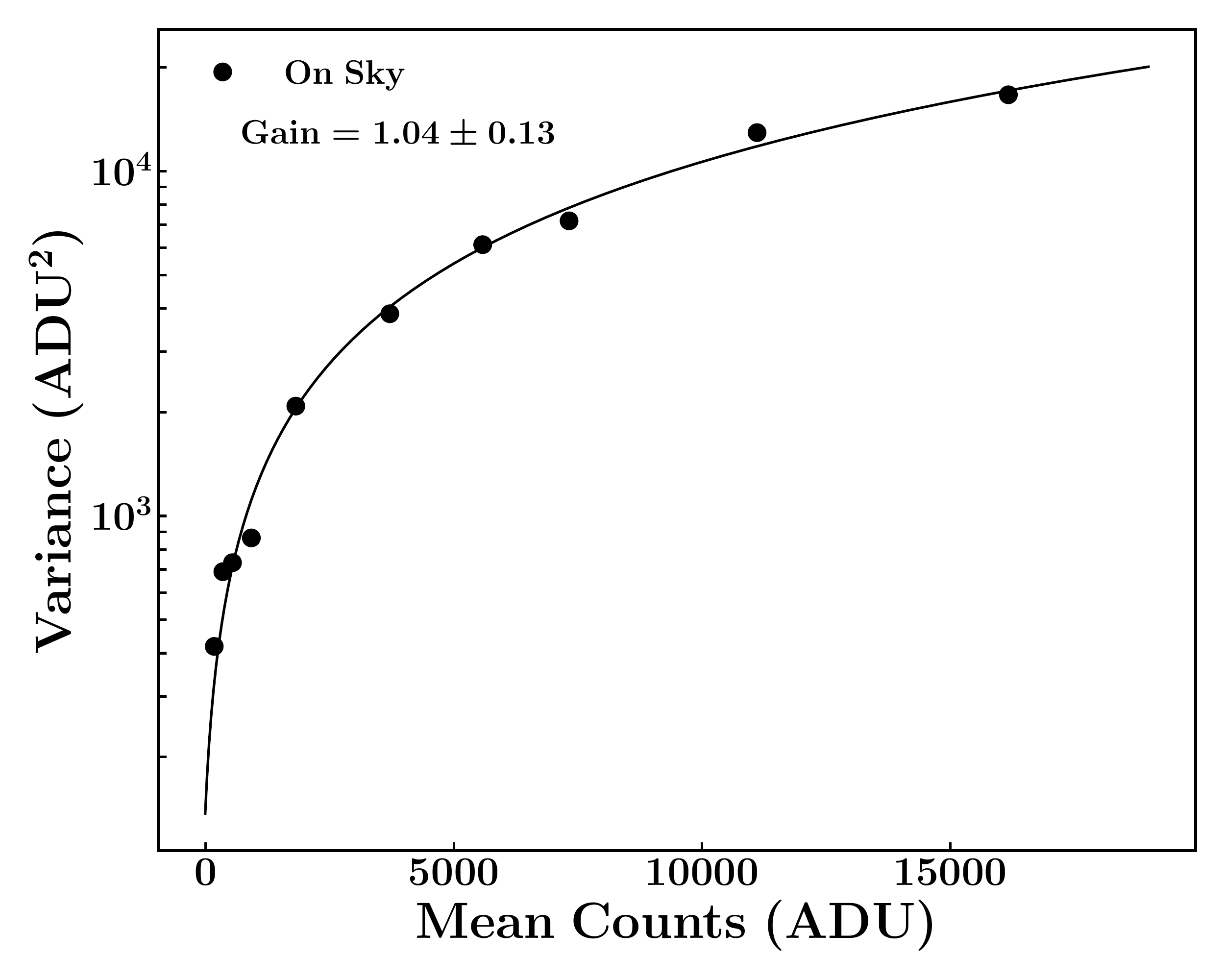}
    \caption{Photon transfer curve (PTC) of the CCD as obtained from the sky experiments. The gain value is estimated as $1.04\pm0.13$ e$^{-}/$ADU.}
    \label{on_sky_gain}
\end{figure}

The estimated gain value using the method described in section \ref{gain} is $1.04\pm0.13$ e$^{-}$/ADU, which is close to the value estimated in the laboratory. On-sky gain estimation is also affected by the sky variation, which results in a slightly higher error bar. The mean-variance plot is shown in Fig. \ref{on_sky_gain}.

\section{Performance of the CCD}
\label{performance}

\begin{figure}
    \centering
    \includegraphics[width=\columnwidth]{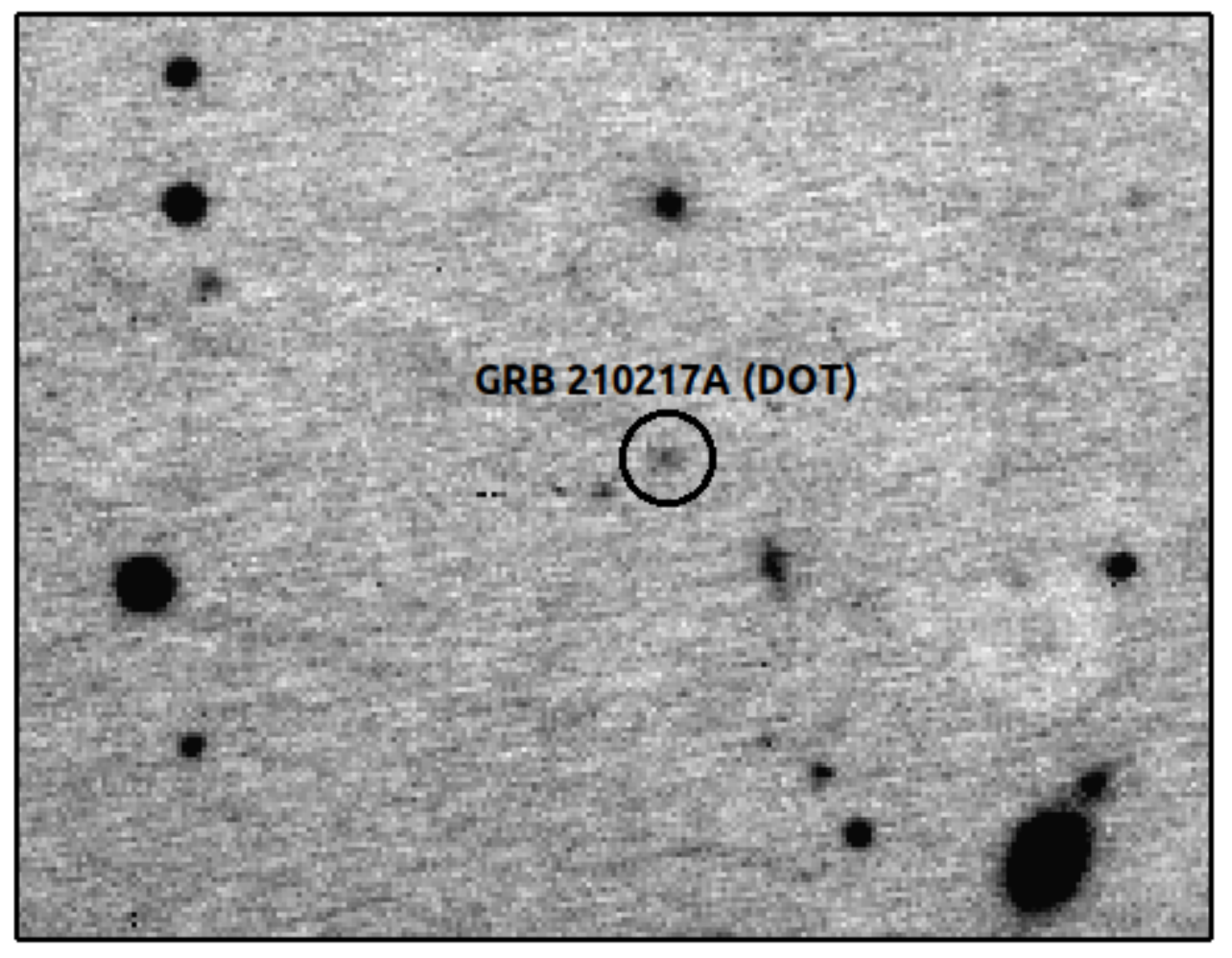}
    \caption{Field of GRB 210217A afterglow imaged with the ADFOSC in r-band.}
    \label{imaging}
\end{figure}

We used the CCD for imaging and spectroscopic observations of various science targets after optimizing it in the lab and successfully verifying it on-sky. This section demonstrates the performance of the CCD system in both imaging and spectroscopic modes with observations of GRB and Supernovae sources. The on-sky performance of ADFOSC on different science targets is discussed in more detail in Omar et al. 2019 \cite{2019arXiv190205857O}.
\subsection{Imaging}
We observed the optical afterglow of GRB 210217A using the ADFOSC in imaging mode. These observations were performed on 18th February 2021 in the r-band at 23:59:18 UT, at $\sim1.7$ days after the burst. Owing to the faintness and rapid decay rate of GRB afterglows, a series of eight images, each with an exposure time of 300 seconds, were acquired. The images were stacked after pre-processing (as described in the previous section) to improve the signal-to-noise ratio. The optical afterglow is visible in the stacked image as shown in Fig. \ref{imaging}. The photometric estimate of the afterglow is $22.32\pm0.16$ mag (AB).

\begin{figure}
     \includegraphics[scale=0.3]{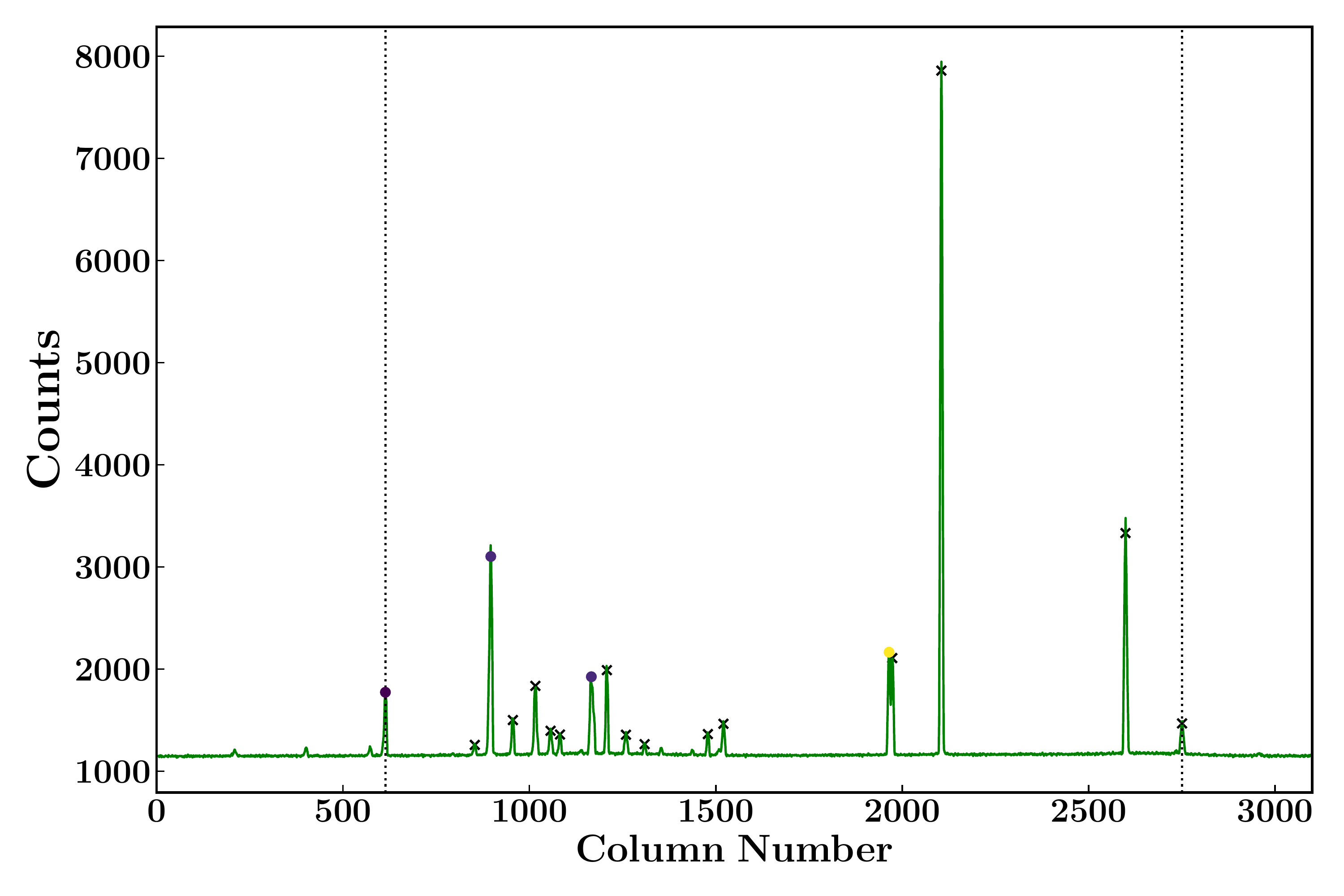}
     \includegraphics[scale=0.3]{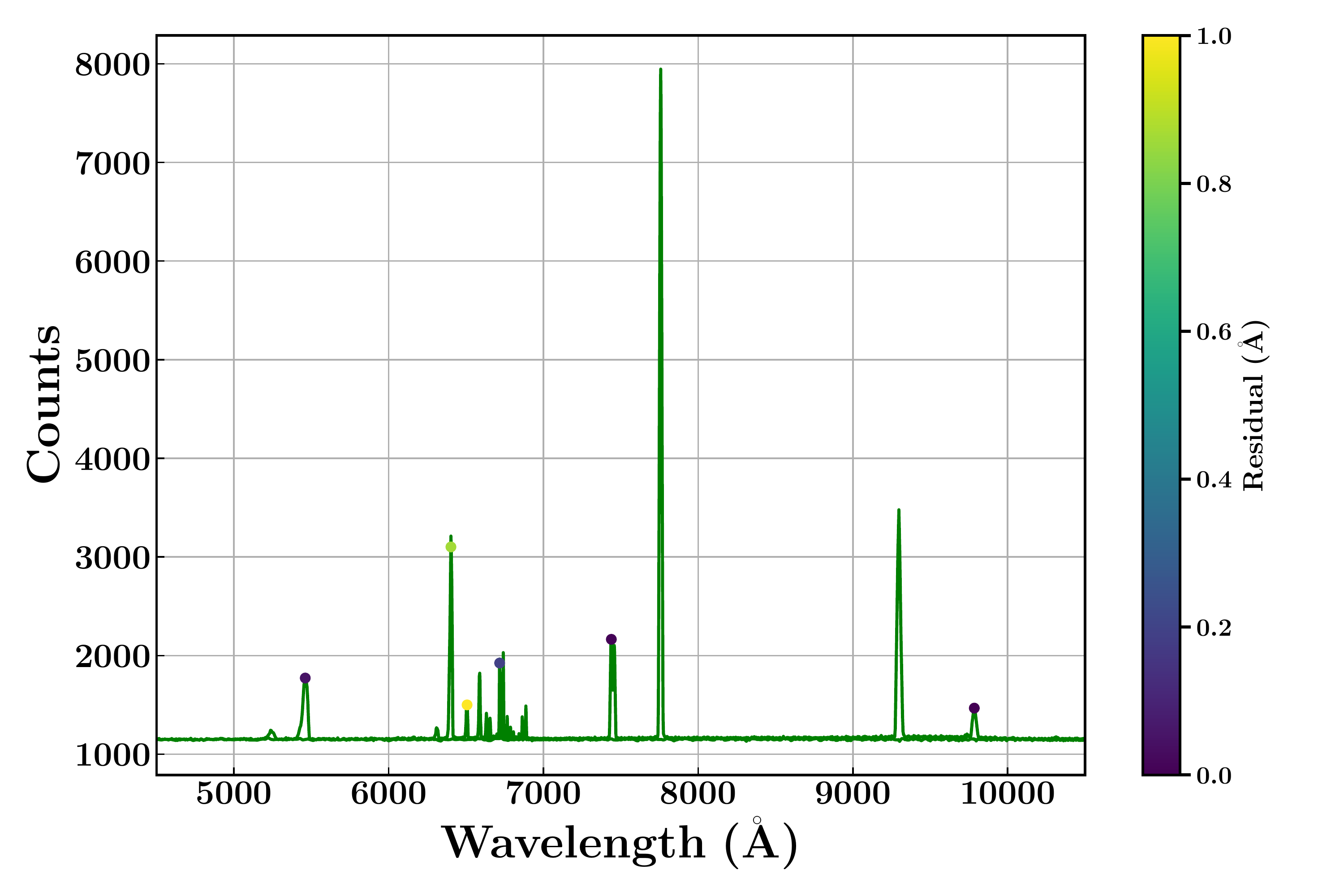}
     \caption{The lamp spectrum of Mercury Argon lamp using 770R grating. The left panel shows the spectrum in pixel scale. The vertical lines indicate the column numbers for the first and last emission lines identified in the spectrum in the left figure. The right panel shows the spectrum in wavelength scale.}
     \label{spectal_dispersion}
\end{figure}

\begin{figure}
    \centering
    \includegraphics[width=\columnwidth]{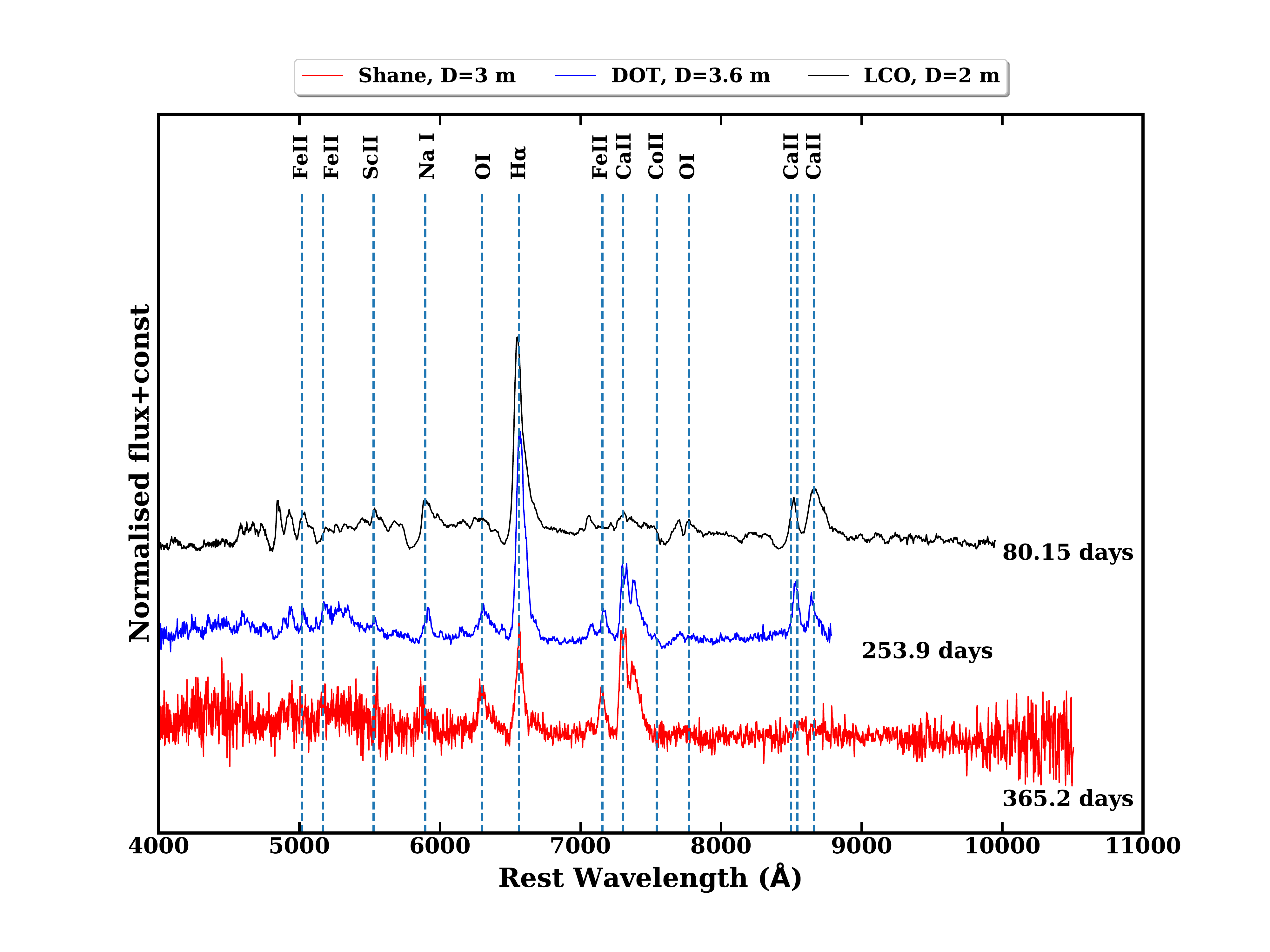}
    \caption{Spectrum of SN~2020jfo obtained using ADFOSC at $\sim254$ days after the discovery of supernova. We identified different absorption lines in the spectrum and compared these with the spectra taken from other instruments/telescopes.}
    \label{spectroscopy}
\end{figure}

\subsection{Spectroscopy}
The spectrograph provides three sets of gratings: 300 gr/mm, 420 gr/mm, and 600 gr/mm. We acquired the lamp spectra using the Mercury Argon (HgAr) calibration lamp to estimate the spectral dispersion. We used the \sw{find peaks} module of \sw{scipy}\cite{2020SciPy-NMeth} to extract the spectral peaks from the obtained spectra. We identified the column number corresponding to each peak and compared it with the wavelength-calibrated lamp spectrum atlas. We defined initial polynomial solutions using these matched wavelength pairs and calculated the best-fit polynomial coefficients to transform between the column number and wavelength. 

The left panel of Fig. \ref{spectal_dispersion} shows the lamp spectrum obtained using a 300gr/nm grating element in pixel scale. The right panel shows the calibrated spectrum in the wavelength scale. A spectral dispersion of 0.20 nm/pixel is estimated for this grating. For gratings elements, 420 gr/mm and 600 gr/mm, the estimated values of spectral dispersion are 0.14 nm/pixel and 0.10 nm/pixel, respectively.

We acquired the spectrum of the supernova SN~2020jfo using $1.5^{''}$ slit and 420 gr/mm grating with an exposure time of 900 sec on 13th January 2021 at 23:12:53 UT. Ailawadhi et al.\cite{2022arXiv221102823A} describe the spectral reduction technique. The absorption features in the spectrum obtained by ADFOSC were identified and matched with the spectra obtained from other telescopes, as shown in Fig. \ref{spectroscopy}. It is noticed that spectral features at longer wavelengths are visible, indicating the high sensitivity of the CCD detector (deep-depleted) in the long-wavelength regime.
  
\section{Conclusions}
\label{conclusions}
We present the methodology employed to characterize the performance of a CCD system developed for integrating with a low dispersion spectrograph instrument, ADFOSC, on the 3.6m DOT. Various experiments were initially performed in the laboratory to characterize and optimize different critical parameters of the CCD system. We also verified the estimated parameters on the sky by mounting the instrument on the 3.6m DOT. We evaluated the bias level during the on-sky tests and examined its stability over several observing nights. We experimentally identified an optimum combination of bias voltages: VOD and VRD, to operate the CCD with minimum non-linearity. The readout performance of the CCD is satisfactory. However, some interference patterns in the image contribute to readout noise. Through different experiments, we tuned and verified the gain parameter corresponding to 1 e$^{-}/$ADU for detecting faint objects. We calculated the dark current at different temperatures using bias frames at lower temperatures and established an optimum operating temperature of the CCD. The CCD acts as a grade-0 detector with no hot pixels at optimum temperature. The regression coefficient values and the gain parameter obtained on-sky are consistent with the values obtained in the laboratory. After verifying the satisfactory performance, we observed the science targets both in imaging and spectroscopic modes. We carried out the imaging of GRB 210217A \cite{Dimple2022} field and the spectroscopy of supernova SN~2020jfo using ADFOSC successfully \cite{2022arXiv221102823A}.
 
\section*{Acknowledgments}
We thank Greg Burley and Tim Hardy from NRC Herzberg Astronomy and Astrophysics Research Centre, Canada, for their help in developing the CCD system. We thank the ARIES 3.6 m DOT engineering team and staff members for providing the necessary support during development, verification, and commissioning work. We would also like to thank Dr. Raya Dastidar for helping us with spectroscopic data reduction.

\bibliography{main}   
\bibliographystyle{spiejour}  

\vspace{1ex}
\section{Biography}
\begin{itemize}
    \item {\bf Dimple} is a PhD student at the ARIES, Nainital. Her research interest mainly focuses on Gamma-Ray Bursts (GRBs) and their associated counterparts: the gravitational waves and the supernovae. She uses multiwavelength data from gamma-ray to optical for her research work.
    \item {\bf Dr. Tripurari S. Kumar} completed his Ph. D. in Systems and Control Engineering from IIT Bombay and serves at ARIES Nainital as head of engineering division primarily working on telescope system engineering aspects. After initially spending two years in Tata Motors Ltd., an automobile industry, he joined ARIES and spent more than eighteen years working on various developmental activities for the ground-based telescopes. His work focuses on development of mechatronics, optoelectronics, motion control and software for telescopes and backend instruments, CCD camera systems, adaptive optics etc. He was lead engineer for development of faint object spectrograph system for the 360-cm Devasthal optical telescope and for assembly integration and verification team of the 360-cm telescope. Currently, he is leading the system engineer aspects of ground-based telescope projects at ARIES which includes upgradation of 360-cm and 130-cm telescopes, development of a new 50 cm space situational awareness telescope jointly with ISRO, adaptive optics for the 360-cm telescope etc. He is one of the members in the India TMT project from ARIES and leading motion control aspects of the WFOS subsystems. 
    \item {\bf Dr. Amitesh Omar} is a scientist working at ARIES, India. His research interests focus on how galaxy evolve in different environments in the Universe. He uses both optical and radio telescopes for his research works. He also takes interests in 'in-country' development of complex back-end instruments for the optical telescopes.
    \item {\bf Dr. Kuntal Misra} is a scientist working at ARIES, India. Her research interests are focused on studying highly energetic transient astrophysical sources and understanding their progenitors. She is also interested in transient search programs using survey data mainly from the 4.0m International Liquid Mirror Telescope (ILMT) located in ARIES. 
\end{itemize}
\listoffigures
\listoftables

\end{spacing}
\end{document}